\documentclass[12pt]{article}
\usepackage{graphicx}
\usepackage{amssymb}
\usepackage{amsfonts}
\usepackage{latexsym}
\usepackage[dvips]{color}

   \newcommand{\beq}{\begin{equation}}
   \newcommand{\eeq}{\end{equation}}
   \newcommand{\be}{\begin{equation}}
   \newcommand{\ee}{\end{equation}}
   \newcommand{\bea}{\begin{eqnarray}}
   \newcommand{\eea}{\end{eqnarray}}

\makeatletter

\@addtoreset{equation}{section}
\makeatother

\def\href#1#2{#2}

\renewcommand{\Re}{\mathrm{Re}\,}
\renewcommand{\Im}{\mathrm{Im}\,}

\begin{document}

\baselineskip=15.5pt
\pagestyle{plain}
\setcounter{page}{1}

\begin{titlepage}
\begin{flushleft}
       \hfill                       FIT HE - 17-01 \\
       \hfill                       
\end{flushleft}

\begin{center}
  {\huge Tunneling in Quantum Cosmology\\   
   \vspace*{2mm}
and Holographic SYM theory 
\vspace*{2mm}
}
\end{center}

\begin{center}

\vspace*{5mm}
{\large ${}^{\dagger}$Kazuo Ghoroku\footnote[1]{\tt gouroku@dontaku.fit.ac.jp},
Yoshimasa Nakano\footnote[2]{\tt ynakano@kyudai.jp},
${}^{\S}$Motoi Tachibana\footnote[3]{\tt motoi@cc.saga-u.ac.jp}\\
and ${}^{\ddagger}$Fumihiko Toyoda\footnote[4]{\tt ftoyoda@fuk.kindai.ac.jp}
}\\

\vspace*{2mm}
{${}^{\dagger}$Fukuoka Institute of Technology, Wajiro, 
Fukuoka 811-0295, Japan\\}
\vspace*{2mm}
{${}^{\S}$Department of Physics, Saga University, Saga 840-8502, Japan\\}
\vspace*{2mm}
{${}^{\ddagger}$Faculty of Humanity-Oriented Science and
Engineering, Kinki University,\\ Iizuka 820-8555, Japan}
\vspace*{3mm}
\end{center}

\begin{center}
{\large Abstract}
\end{center}
We study the time evolution of early universe which is developed by a cosmological constant $\Lambda_4$
and supersymmetric
Yang-Mills (SYM) fields in the Friedmann-Robertson-Walker (FRW) space-time.
The renormalized vacuum expectation value of energy-momentum tensor of the SYM theory 
is obtained
in a holographic way.
It includes 
 a radiation of the SYM field, parametrized as $C$. The evolution is controlled by
this radiation $C$ and the cosmological constant $\Lambda_4$.
For positive $\Lambda_4$, an inflationary solution is obtained at late time. 
When $C$ is added, the quantum mechanical situation at early time is fairly changed.
Here we perform the early time analysis in terms of 
two different approaches, (i) the Wheeler-DeWitt equation and (ii) Lorentzian
path-integral with the Picard-Lefschetz method by introducing an effective action. 
The results of two methods are compared.

\noindent

\begin{flushleft}

\end{flushleft}
\end{titlepage}

\vspace{1cm}


\section{Introduction}

Holographic approach has been extended
to supersymmetric Yang-Mills (SYM) theory  
in Friedmann-Robertson-Walker (FRW) space-time {in Refs.}
\cite{GIN1,EGR,EGR2}. 
There, the vacuum expectation value of the energy-momentum tenser of the SYM fields, 
$\langle T_{\mu\nu}^{\mathrm{SYM}}\rangle$, has been obtained, and it has been used to study
the dynamical properties of the SYM theory in FRW space-time. 
On the other hand, quantum and classical cosmology
with a CFT 
has been studied in terms of this $\langle T_{\mu\nu}^{\mathrm{SYM}}\rangle$\cite{Fisch,GRT,Awad,Barv,Barv2,Bilic}.

{Here, we develop the quantum
cosmology in the system with $\langle T_{\mu\nu}^{\mathrm{SYM}}\rangle$. In both quantum approaches via the path integration and via the Wheeler-DeWitt (WDW) equation, we start
from an action of the theory.}
In the FRW space-time, holographically obtained $\langle T_{\mu\nu}^{\mathrm{SYM}}\rangle$ 
is composed of the loop corrections of the SYM theory
and the so-called dark radiation \footnote{This term has been originally introduced in \cite{BDEL,Lang} and studied
in \cite{SMS,SSM} in the context of the brane universe.}
parametrized by $C$, which
is interpreted as the radiation of SYM fields\cite{GRT}. This stress tensor 
can not be derived
from an general coordinate transformation
invariant action.

Therefore, in performing quantum cosmology with a CFT, we need an effective action which 
can lead to the Einstein equation
including the $\langle T_{\mu\nu}^{\mathrm{SYM}}\rangle$.
Up to now, an example of such 
effective action has been given for the mini-superspace of FRW space-time in Refs. \cite{Fisch,Barv,Barv2}.
In these, however, it is difficult to find the WDW 
equation or to perform the path-integral
due to the high non-linearity.
Here we propose a new and simple effective action, {which reserves the 
essential property of the original theory. 
Then,} it becomes possible to construct the
WDW equation and also to perform a Lorentzian path-integral to study the propagation
of the universe\cite{Turok,Turok3,Turok2}.


We consider here the Einstein gravity with a cosmological constant $\Lambda_4$ and SYM theory (or CFT).
The effective action given here is written
in the same form of the starting action without the SYM theory. Namely, 
after integrating out SYM fields, its effect is reduced to the modification of $\Lambda_4$ to
$\Lambda_4^{\rm eff}$ {which depends on the scale factor $a_0$ and $C$.}
The dark radiation $C$ coming from SYM plays an important role in this effective action
at small $a_0$ of the FRW metric.

{The validity of the Lorentzian path integral has been
shown in \cite{Turok} for the gravity with $\Lambda_4$.
A relevant path in the complex plane of the lapse field could provide a correct propagator of the universe. This propagator
implies an appropriate boundary condition in solving the tunneling amplitude via WDW equation.}
When the dark radiation is absent, $\Lambda_4^{\rm eff}$ is a constant although it is smaller than the 
original $\Lambda_4$. So there is no qualitative change in this case even if we consider CFT.
{As a result, it is also possible to estimate the wave-function of the universe, which is 
created from nothing with no boundary condition, by the Lorentzian path integral via different path 
as shown in \cite{Hartle}.}

On the other hand, when the dark radiation exists, $\Lambda_4^{\rm eff}$ 
is written as a function of $a_0$. Then,
the dynamical situation at early time is drastically changed.
In this case, the scenario is changed as follows:
First, the universe is generated as a small sized sphere of the radiation of SYM fields,
and secondly it reappears as an inflationary universe after the tunneling process. 

Our purpose is to investigate this tunneling behavior by the two methods.
At first, we study by using the WDW equation,
which is derived from our effective action, by imposing an appropriate boundary condition by hand.
In the second, we execute Lorentzian path-integral based on the Picard-Lefschetz theory
to obtain the propagator, as given in \cite{Turok}. 

As shown in \cite{Turok} for the case of the model without SYM theory, namely for $C=0$,
we find that this method
provides the semi-classical tunneling factor which is equivalent with the one obtained by the WDW equation method.
In the case of $C>0$, since 
$\Lambda_4^{\rm eff}$ becomes complicated,
we consider a simplified model to assure this point. 
Then we find the {validity of the Lorentzian path integral method. 
It could give the tunneling amplitude, which is also}
obtained by solving the WDW equation with an appropriate boundary condition.
Other interesting points found in this path integral method are discussed and some speculations are given. 

\vspace{.3cm}
The outline of this paper is as follows.
In the next section, a gravitational model with SYM theory is given 
and the Einstein equations in the FRW space-time are given. They are solved at large $a_0(t)$,
and why quantum cosmology is necessary at small $a_0(t)$ is explained for small $C$.
In Sec. \ref{sec:QCosmology-1}, a tractable effective action corrected by SYM theory is proposed. By using this
action, quantum cosmological solutions at small $a_0(t)$ are shown through the WDW equation in Sec.~\ref{sec:QCos-2}
and through the Lorentzian path-integral method in Sec.~\ref{sec:Lpath-integral}. 
Summary and discussions are given in the final section.


\section{Cosmology driven by CFT}
\label{sec:CbyCFT}

\vspace{.3cm}
Here we consider a model where the matter part is dominated by the ${\cal N}=4$ SYM field with the gauge group $SU(N)$. 
The 4D action is given as
\begin{equation} 
 S=\int d^{4}x\sqrt{-g}\left\{{1\over 2\kappa_4^2}\left(R_{4}-
      {2}\Lambda_4\right)\right\}+S_{\mathrm{SYM}}\,  ,\label{4d-action}
\end{equation}
where $\kappa_4^2\equiv 8\pi G_4$ and $\Lambda_4$ denote the 4D gravitational constant and cosmological constant, respectively.
$S_{\mathrm{SYM}}$ is the action for the ${\cal N}=4$ SYM theory.
After integrating
out all SYM fields 
under the FRW metric (with a scale parameter $a_0(t)$),
the equation of motion for
$a_0(t)$ is obtained by the Einstein equation 
\begin{equation}\label{E2}
  R_{\mu\nu}-{1\over 2}Rg_{\mu\nu}+\Lambda_4g_{\mu\nu}
         =\kappa_4^2 \langle T_{\mu\nu}^{\mathrm{SYM}}\rangle \, ,
\end{equation}
where $\langle T_{\mu\nu}^{\mathrm{SYM}} \rangle$
represents the {vacuum expectation value of the energy momentum tensor for the SYM field
under a given background metric. }

\vspace{.3cm}


Here, Eq. (\ref{E2}) is solved with respect to $a_0(t)$
 under the FRW background
\begin{equation}\label{RW}
  ds_{(4)}^2=-dt^2+a_0(t)^2\gamma_{ij}dx^idx^j\, ,
\end{equation}
where {the 3D} metric $\gamma_{ij}$ is defined as follows:
\begin{equation}\label{AdS4-30} 
    \gamma_{ij}(x)=\delta_{ij} \gamma^2(x)\, , \quad \gamma (x)
  =\frac{1}{ 1+k{\bar{r}^2\over 4\bar{r}_0{}^2}}\, , \quad 
    \bar{r}^2=\sum_{i=1}^3 (x^i)^2\, .
\end{equation}
Then two independent equations are obtained such that
\begin{equation}\label{bc-RW2}
 \lambda\equiv  \left({\dot{a}_0\over a_0}\right)^2+{k\over a_0^2} = {\Lambda_4\over 3}+{\kappa_4^2\over 3} 
\langle T_{00}^{\mathrm{SYM}}\rangle\, ,
\end{equation}
\begin{equation}\label{bc-RW3}
 2{\ddot{a}_0\over a_0} +\left({\dot{a}_0\over a_0}\right)^2+{k\over a_0^2} =  {\Lambda_4}-{\kappa_4^2\langle T_{ii}^{\mathrm{SYM}}\rangle} \, .
\end{equation}
Eq. (\ref{bc-RW2}) {is nothing but} the $tt$ component of the Einstein equation,
  \textit{i.e.\/}, the Friedmann equation.

From (\ref{bc-RW2}) and (\ref{bc-RW3}), we obtain the following continuity equation 
for density $\rho$ and pressure $p$ of the SYM fields\cite{GN13,KSS,BFS,FG}:
\begin{equation}
 \dot{\rho}+3H(\rho +p)=0\, , \quad  H\equiv \frac{\dot{a}_0}{a_0} ,
 \label{continuity-1}
\end{equation}
where the averaged energy-momentum tensor is written in terms of $\rho$ and $p$ as
\begin{equation}
  \langle T_{\mu\nu}^{\rm SYM} \rangle ={\rm diag}(\rho, pg_{ij}^0)\, ,
\end{equation}
and
\begin{eqnarray}
 \rho&=&3 \alpha \left(\frac{C}{4R^2a_0^4}+\frac{\lambda^2}{16}\right) \, ,
\quad
 p=\alpha\left\{ \frac{C}{4R^2a_0^4}-3 {\lambda^2\over 16}\left(
     1+{2\dot{\lambda}\over 3\lambda}{a_0\over \dot{a}_0}\right) \right\}\, ,
\label{eq:density_pressure}\\
&&\hspace{1.5in}\alpha={4R^3\over 16\pi G_N^{(5)}}\, ,
\label{eq:alpha}
\end{eqnarray}
{where $R$ denotes the radius of the AdS$_5$,} and
$C$ being the dark radiation density, regarded as the radiation of CFT. 
{ $G_N^{(5)}$ denotes the 5D gravitational constant,
and $g_{ij}^0=a_0(t)^2\gamma_{ij}$ \cite{GN13}. }

Note here that solving
 (\ref{bc-RW2}) and (\ref{bc-RW3}) is equivalent to solving
  (\ref{bc-RW2}) and (\ref{continuity-1})
since (\ref{continuity-1}) is derived from (\ref{bc-RW2}) and (\ref{bc-RW3}).
On the other hand,  
(\ref{continuity-1}) is satisfied for $\langle T_{\mu\nu}^{\mathrm{SYM}}\rangle$.
Therefore, it is enough to solve Eq. (\ref{bc-RW2}) to obtain $a_0(t)$.

\vspace{.3cm}
{Here, Eq. (\ref{bc-RW2})
is written by using $\lambda$ as}
\begin{equation}
  \lambda 
={\Lambda_4\over 3}+\tilde{\alpha}^2\left(\frac{4C}{R^2a_0^4}
+\lambda^2 \right)\, ,\label{Freed-2}
\end{equation}
where
\begin{equation}
 \tilde{\alpha}^2={\kappa^2_4\over 16}\alpha\, .
\end{equation}
Then (\ref{Freed-2}) is solved with respect to $\lambda$ as 
\begin{equation}\label{a0-eq}
  \lambda=\lambda_{\pm}\label{eq:lambda=lambda+-}
\end{equation}
where
\begin{equation}
\lambda_{\pm}\equiv {1\pm\sqrt{1-4\tilde{\alpha}^2
   \left(\frac{\Lambda_4}{3}+\frac{4\tilde{\alpha}^2C}{R^2a_0^4}\right)}\over 2\tilde{\alpha}^2}\ .\label{eq:lambda+-}
\end{equation}
The explicit form of (\ref{a0-eq}) is given as\footnote{
 We remember that
$\displaystyle
\left({\dot{a}_0\over a_0}\right)^2+{k\over a_0^2}=\lambda\, . \label{bc-3-1}
$}
\begin{equation}
     \left({\dot{a}_0\over a_0}\right)^2+{k\over a_0^2}
 = {1\pm\sqrt{1-4\tilde{\alpha}^2
   \left(\frac{\Lambda_4}{3}+\frac{4\tilde{\alpha}^2C}{R^2a_0^4}\right)}\over 2\tilde{\alpha}^2}\, .\label{Fried-3}
\end{equation}

\vspace{.3cm}
When we solve
(\ref{Fried-3}), we must notice the following points.
At first for finite $C$, from the reality of this equation,
we find that there is a minimum value of $a_0$ such as
\begin{equation}\label{constraint}
  a_0 \geq  a_0^\mathrm{min}=\tilde{\alpha} \left({16C\over R^2\tilde{\Lambda}_4}\right)^{1/4}\, ,
     \quad   \tilde{\Lambda}_4=1-4\tilde{\alpha}^2 \frac{\Lambda_4}{3} .
\end{equation}
For the case with $a_0< a_0^\mathrm{min}$, we need some improvement of the gravitational theory. 
This point remains as an open problem here.
While the solutions $a_0(t)$ for $\lambda_{+}$ and $\lambda_{-}$ can be connected at $a_0=a_0^\mathrm{min}$\cite{Awad},
we consider here only the case with $\lambda_-$. {The reason of this choice is as follows: 
In the limit $\tilde{\alpha}\to 0$, it would be natural
that $\lambda \to \Lambda_4/3$. While $\lambda_+$ diverges, $\lambda_-$ approaches to $\Lambda_4/3$ in this limit.} 

In the case of $C=0$, the above constraint (\ref{constraint}) leads to the upper bound of $\Lambda_4$,
\begin{equation}
\Lambda_4\leq {3 \over 4\tilde{\alpha}^2}\, .
\end{equation} 
This is an interesting result given as the quantum effect of the SYM theory since 
we need some physical reason to suppress the cosmological constant to probably zero.
{Since $1/\tilde{\alpha}^2\propto 1/R^3$, it is possible to control $\Lambda_4$ by an appropriate choice
of the radius $R$ of AdS$_5$.}

\vspace{.3cm}
It is easy to find a classical solution corresponding to $\lambda_{-}$. From (\ref{Fried-3}), 
the equation to be solved is given by
\begin{equation}\label{Fried-4}
\dot{a}_0^2=-k+a_0^2\frac{1-\sqrt{1-4\tilde{\alpha}^2(\frac{\Lambda_4}{3}+\frac{4\tilde{\alpha}^2C}{R^2a_0^4})}}
{2\tilde{\alpha}^2} \ \ (\equiv-2V(a_0))
\end{equation}
Here we consider the case of $\Lambda_4>0$ and $k=1$ (closed universe).
At large $a_0$, we always find the inflationary solution, \textit{i.e.\/},
\begin{equation}
  {a}_0(t)\sim \exp\left(\gamma~t \right)\, , 
\quad \gamma = { \sqrt{\frac{1-\sqrt{1-4\tilde{\alpha}^2\frac{\Lambda_4}{3}}}
{2\tilde{\alpha}^2}}}\ .
\end{equation} 
Note that the expansion rate or the effective cosmological constant is suppressed by the quantum effect of the SYM theory.  
{For small $\tilde{\alpha}$ or for small number of SYM fields, the factor $\gamma$ approaches
to $\sqrt{\Lambda_4/3}$ as expected.}

\vspace{.3cm}
On the other hand, in the small $a_0$ region, we should be careful of 
the potential $V(a_0)$ defined in (\ref{Fried-4}).
 Let us classify this situation into the following three types.

\begin{figure}[htbp]
\vspace{.3cm}
\begin{center}
\includegraphics[width=10.0cm,height=7cm]{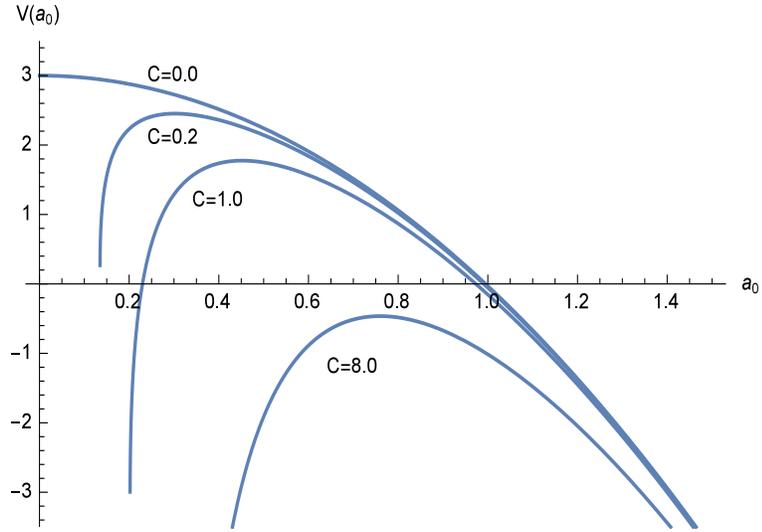}
\caption{Effective potential for $R=1.0, ~\Lambda_4/3=1.0, ~k=1,~\tilde{\alpha}^2=0.01$.}
\label{W-D-1}
\end{center}
\end{figure}

\noindent
(a) For $0\leq C\leq \frac{1}{4}k^2R^2(1-4\tilde{\alpha}^2\frac{\Lambda_4}{3})$,
the classical solution is restricted to the region $a_0^{+}<a_0$, where $V(a_0^+)=0$.
Such a behavior is seen for the parameter region $0 \leq C \leq 0.24$ in Fig.~\ref{W-D-1}.
\footnote{Hereafter, we use $R=1.0$ for every numerical estimation, so the value of $R$ is not denoted except for
a special case.}

\begin{figure}[htbp]
\vspace{.3cm}
\begin{center}
\includegraphics[width=7.5cm]{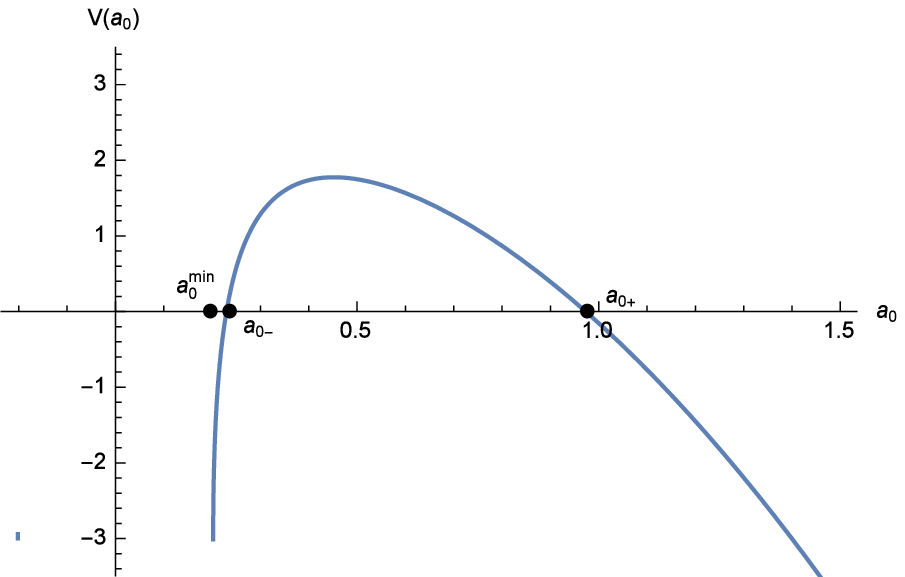}
\includegraphics[width=7.5cm]{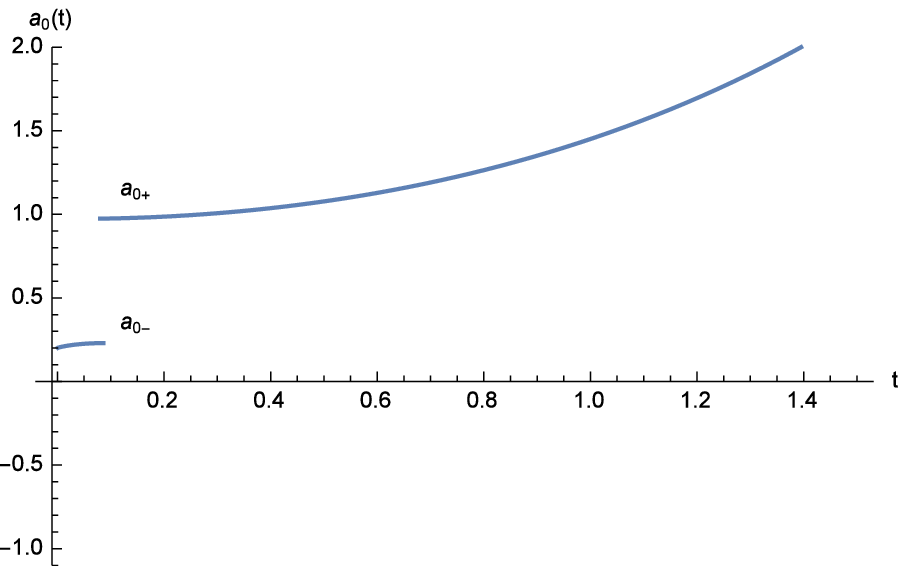}
\caption{Left: Potential barrier of $C=1.0$.
Right: Two independent classical solutions for $a_0(t)$.}
\label{W-D-2}
\end{center}
\end{figure}

\noindent
(b) For $ \frac{1}{4}k^2R^2(1-4\tilde{\alpha}^2\frac{\Lambda_4}{3}) \leq C \leq  
\frac{16\tilde{\alpha}^2\Lambda_4}{3}k^2R^2(1-4\tilde{\alpha}^2\frac{\Lambda_4}{3})$,
there appears the potential barrier in the region $a_{0-}\leq a_0 \leq a_{0+}$ (see the left panel of Fig.~\ref{W-D-2}).
 In this region, no classical solutions are allowed.


In the region $a_0^\mathrm{min} <a_0< a_{0-}$ (see the left panel of Fig.~\ref{W-D-2} again), 
there is a classical solution, but the solution does not provide the inflationary one.
On the other hand, in the region  $a_0>a_{0+}$, we obtain the inflationary solution.
These two solutions are not connected to each other at the classical level.
But they can be connected at the quantum level through the tunneling effect.
 See the right hand figure of Fig.~\ref{W-D-2}.
In \cite{Barv}, a new hilltop inflation scenario has been studied 
by using the Euclidean time solution (instanton) for this region.
 
\noindent
(c) For $C \geq \frac{16\tilde{\alpha}^2\Lambda_4}{3}k^2R^2(1-4\tilde{\alpha}^2\frac{\Lambda_4}{3})$,
there is no potential barrier and the classical inflationary solution is obtained for $a_0^{\rm min}<a_0$.

In the following sections, we consider the time evolution of the universe in the small $a_0$ region
by using quantum cosmological approaches.


\section{Quantum Cosmology and Effective Action}
\label{sec:QCosmology-1}

There are two quantum mechanical ways to study the small $a_0$ region,
canonical approach and path-integral approach.{ In any case, we need an effective action
which leads to the classical equations of motion (\ref{E2}) for an appropriate coordinate.
It would be impossible to find a general coordinate invariant form since $\langle T_{\mu\nu}^{\mathrm{SYM}}\rangle $
comes from the conformal anomaly for the CFT.}

{However, we could find an effective action which provide the classical equations of motion, (\ref{bc-RW2}) and 
(\ref{bc-RW3}), written in the FRW metric. Namely, when the dynamical variable is restricted
to the scale factor with a lapse function as a multiplier, it becomes possible to extend the analysis to
the quantum cosmology.}

Then we restrict ourselves in mini-superspace.
Let us consider the metric
\begin{equation}\label{RW-N}
  ds_{(4)}^2=-{\cal N}^2(t)dt^2+a_0(t)^2\gamma_{ij}dx^idx^j\, ,
\end{equation}
where ${\cal N}(t)$ denotes the lapse function.
According to the idea of {\cite{Fisch,Barv2},}
the effective action is written as follows:
\footnote{Here an appropriate boundary term is abbreviated.} 
\begin{eqnarray} 
 S&=&\int\!\!d^{4}x\sqrt{-g}\left\{{1\over 2\kappa_4^2}\left(R_{4}-
      {2}\Lambda_4\right)+L_{\mathrm{SYM}}^{\mathrm{eff}}\right\}\,  \nonumber \label{4d-action-q1} \\
  &=&V_3\!\!\int\!\!dt\,{\cal N}a_0^3 \left\{{1\over \kappa_4^2}
      \left({3\over a_0^2}(-{\dot{a}_0^2\over {\cal N}^2}+k)-
      \Lambda_4\right)+L_{\mathrm{SYM}}^{\mathrm{eff}}\right\}\, , \label{4d-action-q2}    
\end{eqnarray}
where 
\begin{equation}\label{effL-2}
    L_{\mathrm{SYM}}^{\mathrm{eff}}
=-{3N^2\over 32\pi^2}\left(\frac{4C}{R^2a_0^4}+{k^2\over a_0^4}
        -{2k\over a_0^2}{\dot{a}_0^2\over {\cal N}^2a_0^2}-{1\over 3{\cal N}^4}{\dot{a}_0^4\over a_0^4}\right)\, .\label{eq:LeffSYM}
\end{equation}

\vspace{.3cm}
The Lagrangian $L_{\mathrm{SYM}}^{\mathrm{eff}}$ 
is determined such that we could
find the Friedmann equation (\ref{bc-RW2}) from the stationary condition
for ${\cal N}$
with ${\cal N}=1$ gauge. We should notice that
(\ref{bc-RW3}) is also found from the variational equation of $a_0$. In this sense, 
(\ref{4d-action-q2}) with (\ref{effL-2})
leads to a correct form of equations of motion to obtain our classical solutions.
It is however difficult 
to develop an effective quantum theory based
upon the action (\ref{4d-action-q1}) due to the term (\ref{effL-2}).
The situation is similar to the case with higher curvature terms.

Then we consider an alternative effective action which leads to (\ref{Fried-3}) instead of (\ref{bc-RW2}).
It is given as
\begin{eqnarray} 
 S &=& \int\!\!d^{4}x\sqrt{-g}{1\over 2\kappa_4^2}\left(R_{4}-2\Lambda_{\rm eff} \right)\,
   \nonumber \label{4d-action-q3}  \\
  &=& V_3\!\!\int\!\!dt\,{\cal N}a_0^3 {1\over \kappa_4^2}
      \left({3\over a_0^2}(-{\dot{a}_0^2\over {\cal N}^2}+k)-
      \Lambda_{\rm eff}\right)\,  ,\label{4d-action-q4}    
\end{eqnarray}
where
\begin{equation}\label{Lam-eff}
 \Lambda_{\rm eff}^{\pm}
=3\lambda_{\pm}
\equiv 3\,{1\pm\sqrt{1-4\tilde{\alpha}^2
 \left(\frac{\Lambda_4}{3}+\tilde{\alpha}^2\frac{4}{R^2}\frac{C}{a_0^4}\right)}\over 2\tilde{\alpha}^2}\ .
\end{equation}
This action is useful to perform the canonical formulation. In fact,
from this action, it is easy to obtain the WDW equation
 (see Appendix \ref{sec:WDWequation}) which is not written here
since we change the variable from $a_0$ to $q=a_0^2$.

\vspace{.3cm}
\subsection{Change of Variables}

According to {\cite{Turok,Turok3}}, $a_0$ and ${\cal N}$ are changed as $q= a_0^2$ and ${\cal N}\rightarrow{\cal N}/a_0$,
then we have
\begin{eqnarray}
  S&=& {V_3 \over \kappa_4^2}\int\!\!dt \left( -{3\dot{q}^2\over 4\mathcal{N}}+\mathcal{N}(3k-q\Lambda_{\mathrm{eff}}) \right)\, \label{action-2}\ ,\\
  \Lambda_{\mathrm{eff}} &=& 3\,
\frac{~~1-\sqrt{1-4\tilde{\alpha}^2\left(
\frac{\Lambda_4}{3}+\tilde{\alpha}^2\frac{4}{R^2}\frac{C}{q^2}\right)}~~}{2\tilde{\alpha}^2}\ .
\end{eqnarray}
In this formulation, the WDW equation is given as 
\begin{equation}
\left(-\hbar^2\frac{\partial^2}{\partial q^2}
+V(q)
\right)\,\Psi(q)=0\ ,\label{WDW-eq-2}
\end{equation}
which is also obtained by changing the variable as $q=a_0^2$ in (\ref{WDW-eq}),
The potential $V(q)$ is given by
\begin{equation}
  V_{\mathrm{eff}}(q)\equiv V(q)/(3v_3^2)=3k-q \Lambda_{\mathrm{eff}}(q)\ .
\label{Veff-rev}
\end{equation}

\subsection{$V_{\rm eff}(q)$ and tunneling}

As shown in Sec.~\ref{sec:CbyCFT}, 
we find similar behavior of the potential $V_{\rm eff}(q)$ to the case of $V(a_0)$.
The typical potentials are similar to Fig.~\ref{W-D-1}, so they are abbreviated here.

\vspace{.3cm}
Here we should notice as mentioned in Sec.~\ref{sec:CbyCFT} that there is a lower bound of $q$ for $C>0$. 
It corresponds to $a_0^{\rm nin}$ given in the previous section. The bound is given as
\begin{equation}
  q^\mathrm{min}
   =\tilde{\alpha}^2 \left({16C\over R^2\tilde{\Lambda}_4}\right)^{1/2}\, ,
\quad
 \tilde{\Lambda}_4=1-\frac{4\tilde{\alpha}^2 \Lambda_4}{3}\, .
\label{q-min}
\end{equation}
For $q<q^\mathrm{min}$, the potential becomes complex.
So we cannot extend our model in this region. 
Therefore we concentrate
our analysis on the region $q>q^\mathrm{min}$. 

\begin{figure}[htbp]
\vspace{.3cm}
\begin{center}
\includegraphics[width=10.0cm,height=7cm]{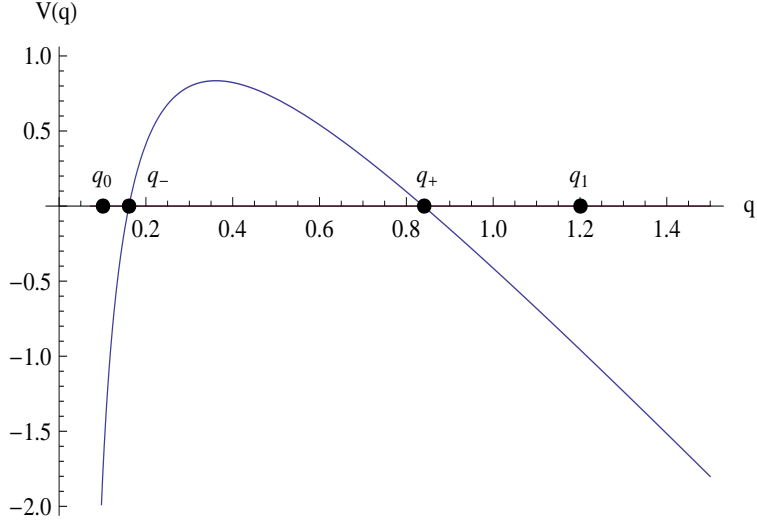}
\caption{Plots of $V_{\mathrm{eff}}(q)$ vs $q$ for $\Lambda_4=3, ~k=1,~\tilde{\alpha}^2=0.01$ and
 $C=31.3$.
\label{Veff-q-dS}}
\end{center}
\end{figure}

In this allowed region, several types of potentials are seen depending on the value of $C$.
Hereafter we study the case shown in Fig.~\ref{Veff-q-dS}. 
This has two turning points, say $q_-$, and $q_+(>q_-)$,
in the two classical regions.
For very small $C$, small sized universes may be made but they will soon disappear into the 
region of $q<q^\mathrm{min}$ where Einstein-Hilbert action is not available. 
However some of them would go through the mountain of the potential
via a quantum tunneling effect. From the viewpoint of inflational scenario, this quantum jump
of a small sized universe would give us a clue to the initial condition of the inflation.

Then, by using this potential,
the tunneling birth of our universe can be studied.

\vspace{.2cm}
In principle, it is possible to calculate the propagator, $G(q_1,q_0)$ where the points $q_0$ and $q_1$
are shown in Fig.~\ref{Veff-q-dS}, according to the path-integral as discussed above
in order to see the tunneling effect. However, it is difficult to find a saddle points in the complex
${\cal N}$ plane in the present model. So we perform the same calculation in terms of solving
the WDW equation as given follows.


\section{WDW equation and tunneling}
\label{sec:QCos-2}


Considering the potential as shown in Fig.~\ref{Veff-q-dS},
we give the tunneling amplitude for $C>0$ by solving the WDW equation (\ref{WDW-eq-2}). 
Supposing the form of the wave-function $\Psi(q)$ as follows:
\begin{equation}
  \Psi(q)=A(q) e^{i\phi(q)/\hbar}\,  ,
\end{equation}
the WDW equation (\ref{WDW-eq-2}) leads to the following equations,
\begin{eqnarray}
  (\partial_q\phi)^2+12\pi^4{V_{\rm eff}}&=& {\hbar^2\over A}\partial_q^2 A
\, , \label{WKB-1}\\
  \partial_q A\partial_q\phi +{1\over 2} A \partial_q^2 \phi &=&0 \, . \label{WKB-2}
\end{eqnarray}
They are solved by expanding $\phi$ with the power of $\hbar$.

In the regions of $q<q_-$ and $q_+<q$, the wave functions $\Psi_1$ and $\Psi_3$ are obtained in the 
following forms
\begin{eqnarray}
  \Psi_1&=&{c_+\over \sqrt{k(q)}} e^{-i\eta(q,~q_-)}+{c_-\over \sqrt{k(q)}} e^{i\eta(q,~q_-)}\, , \\
  \Psi_3&=&{D_+\over \sqrt{k(q)}} e^{-i\eta(q_+,~q)}+{D_-\over \sqrt{k(q)}} e^{i\eta(q_+,~q)}\, ,
\end{eqnarray}
where {both the first terms of $\Psi_1$ and $\Psi_3$ 
 represent} the outgoing (growing) wave, and  
\begin{equation}\label{WDW-pot-1}
   k(q)=\sqrt{|V|}/\hbar\, , \quad \eta(q,~q_-)=\int_q^{q_-} dq' k(q')\, ,
\end{equation}
where we notice that $ \eta(q,~q_-)=P(q,~q_-)$ which is defined in the next section
to express the
Green function $G(q,~q_-)$.

We can set various boundary conditions for the solutions of the wave-function $\Psi_q$.
At first, we concentrate on the tunneling.
In order to see the tunneling amplitude,
we impose the condition $D_-=0$. Namely only the outgoing wave is restricted in the region
$q_+<q$, then we find 
\begin{equation}
  D_+=c_+/\left({1\over 4} e^{-\eta(q_-,~q_+)}+ e^{\eta(q_-,~q_+)}\right)\ .
  \end{equation}
Then we find the tunneling probability
\begin{equation}
  T\simeq \left|{D_+\over c_+} \right|^2=e^{-2P(q_+,q_-)}\, .
  \label{WKB-T}
\end{equation} 

\begin{figure}[htbp]
\vspace{.3cm}
\begin{center}
\includegraphics[width=10.0cm,height=7cm]{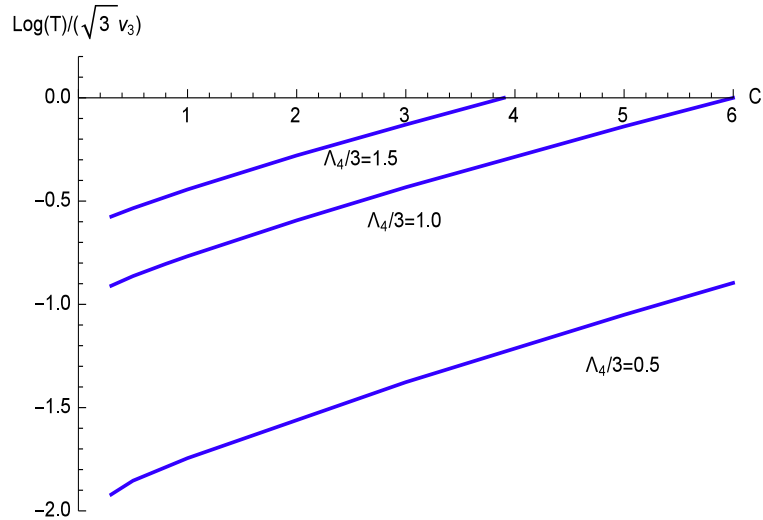}
\caption{Large $C$ dependence of the tunneling probability $T$ for different values of $\Lambda_4$}
\label{fig:WKB-T}
\end{center}
\end{figure}

We expect that this result could also be found from the propagator given by the path-integral discussed above\cite{Turok}.

On the other hand, for the condition $D_+=-iD_-$, we find the relation
\begin{equation}
    c_+={D_+\over 2}e^{-\eta(q_-,~q_+)}\, .
\end{equation}
This leads to
\begin{equation}
   \left|{D_+\over c_+} \right|^2=4\,e^{2P(q_+,q_-)}\, .
\end{equation}
The sign of the exponent of $ \left|{D_+\over c_+} \right|^2$ in this case is opposite to the tunneling case.
This result corresponds to the one obtained in \cite{Hartle} as the wave-function of the WDW equation.

There are many other conditions, which lead to various forms of $ \left|{D_+\over c_+} \right|^2$. The point we want to see is
how these solutions of the WDW equation are related to the results of the path-integral. The saddle
points given from the effective action of complex ${\cal N}$ can be related to the above solutions of  the WDW
equation. In order to answer to this question, we consider a simple model which has the properties of the
above holographic model for $C>0$ case. 


\section{Quantum cosmology with Lorentzian path-integral}
\label{sec:Lpath-integral}

In the previous section, we considered canonical formalism
to find the wave function of the universe
by the WDW equation. In this section, let us consider the path-integral formalism
to get the propagator of the universe. 

The Feynman propagator in mini-superspace is defined as \cite{Turok}
\begin{equation}
  G( q_1, q_0)=\int_{0^+}^{\infty} d{\cal N} \int_{q_0}^{q_1} Dq e^{iS({\cal N},q)/\hbar }\, ,\label{eq:propagatorG}
\end{equation}
where $q_0=q(0)$ and $q_1=q(t_1)$ are the initial and final values of $q(t)$. 
After integrating over $q$, we are left with the integration over ${\cal N}$.
In order to perform the integration, we here try to apply the Lefschetz thimble method.
In the Appendix C, the detail of this method is shown.

In this method, the original path is extended to the complex plane,
\begin{equation}
   {\cal N}=u_1+iu_2\ ,
\end{equation}
where both $u_1=\Re({\cal N})$ and $u_2=\Im({\cal N})$ are real,
and the propagator is evaluated by the saddle point
approximation in the $\hbar \rightarrow 0$ limit. Below we shall see those examples concretely.

\vspace{.3cm}
\subsection{The case of $C=0$}
In this case, the action (\ref{4d-action-q4}) is equivalent to the one with the Einstein-Hilbert term and
a cosmological constant. So, it has been already studied in \cite{Turok}.  Equation of motion for $q(t)$ and the constraint from (\ref{action-2}) are
\begin{eqnarray}
  \ddot{q}&=&{2\over 3}\mathcal{N}^2 \Lambda_{\mathrm{eff}} , \label{q-eq}\\
  {3\dot{q}^2\over 4\mathcal{N}^2}+3k&=&q \Lambda_{\mathrm{eff}}\ .
\end{eqnarray}

The path integral over $q$ becomes Gaussian and is exactly treated. Then we are left with the integral
over ${\cal N}$ as follows:
\begin{eqnarray}
G( q_1, q_0)= \sqrt{\frac{3\pi i}{2\hbar}}\int_0^{\infty} \frac{d{\cal N}}{{\cal N}^{1/2}}e^{2\pi^2 i S({\cal N})},
\label{N-int}
\end{eqnarray}
where
\begin{eqnarray}
S({\cal N})=\frac{1}{36}{\cal N}^2\Lambda_{\mathrm{eff}}^2+\left( 3k-\frac{1}{2}(q_0+q_1)\right ){\cal N}
-\frac{3}{4{\cal N}}(q_1-q_0)^2.
\label{S_N}
\end{eqnarray}

\begin{figure}[htbp]
\vspace{.3cm}
\begin{center}
\includegraphics[width=9cm,height=9cm]{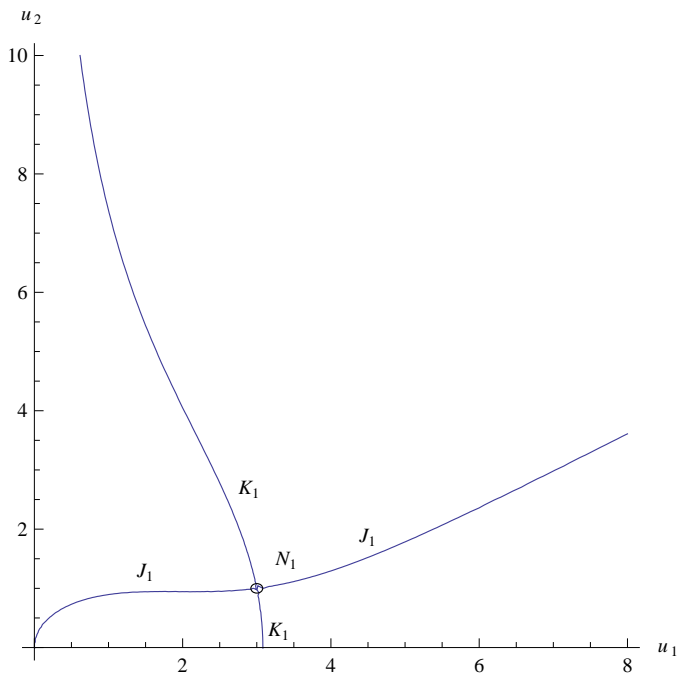}
\caption{Plots of Lefschetz thimble $J_1$ (the steepest descent) and $K_1$ (the steepest ascent) for the saddle $\mathcal{N}_1=(3,~i)$, where
$\Lambda_{\mathrm{eff}}=3$, $k=1$, $q_0=0$, $q_1=10$. 
\label{Lefschetz-thimble}}
\end{center}
\end{figure}

The action
(\ref{S_N}) has four saddle points in the complex ${\cal N}$ plane. If we choose, for instance,
$k=1, \ \Lambda_{\mathrm{eff}}=3, \ q_0=0$ and $q_1=10$, those points lie at $(3,\ i), \ (3, -i), \
(-3, i)$ and $(-3, -i)$. Then, by using the thimble decomposition (as described in the Appendix C), 
it is found that the original contour is deformed to the Lefschetz thimble $J_1$
so that only one saddle ${\cal N}_1=(3, \ i)$ contributes to the integral (see Fig.~\ref{Lefschetz-thimble}).


The propagator obtained in this way 
is interpreted as
the tunneling probability factor when the initial value $q_0$ is
in the quantum region and the final point is at the zero point of $V_{\rm eff}$. It is given by
\begin{equation}\label{tunnelG}
   G( q_1, q_0)\propto e^{-2 P(q_1,q_0)}\, , 
\end{equation}
where 
\begin{equation}\label{tunnel}
 P(q_1,q_0)
  = {v_3 \over \hbar}\int_{q_0}^{q_1}\!\!\sqrt{3V_{\rm eff}^{(0)}(q)}\,dq\ ,  
\quad  V_{\rm eff}^{(0)}(q)=3k-q \Lambda_{\mathrm{eff}}^{(C=0)}(q)\, .
\end{equation}
Here the oscillating part is abbreviated.
This term precisely denotes  the 
tunneling probability found in the WKB approximation of the WDW equation in the previous section.

Here is a comment. In this case, we find
other three saddle points 
which do not contribute to the integration. However, they might affect in some case
where the integration path is defined in a different way\cite{Hartle}. 

The tunneling factor (\ref{tunnel}) appears whenever either $q_0$ or $q_1$ is in the quantum region.
So the Lefschetz thimble method is useful to study quantum cosmology and it is equivalent to solve the WDW equation
 under appropriate boundary conditions.

\subsection{The case of $C\neq 0$}

It is difficult to perform the path-integral for the potential $V_{\rm eff}(q)$
with $C\neq 0$ due to a complicated $q$-dependence of $V_{\rm eff}(q)$.
Then let us consider the following action, that is,
\begin{eqnarray}
  S&=& {v_3}\!\int\!dt\left( -{3\dot{q}^2\over 4{\cal N}}
  +{\cal N}(3k-\mu (q_c -q)^2-\beta) \right)\, \label{action-2-2} \qquad
(\mu >0)\ .  \label{simple}
\end{eqnarray}
Here the original $V_{\mathrm{eff}}(q)$ defined by (\ref{Veff-rev})
is approximated by a quadratic form 
near its maximum point, $q_c$. 

\begin{figure}[htbp]
\vspace{.3cm}
\begin{center}
\includegraphics[width=10.0cm,height=7cm]{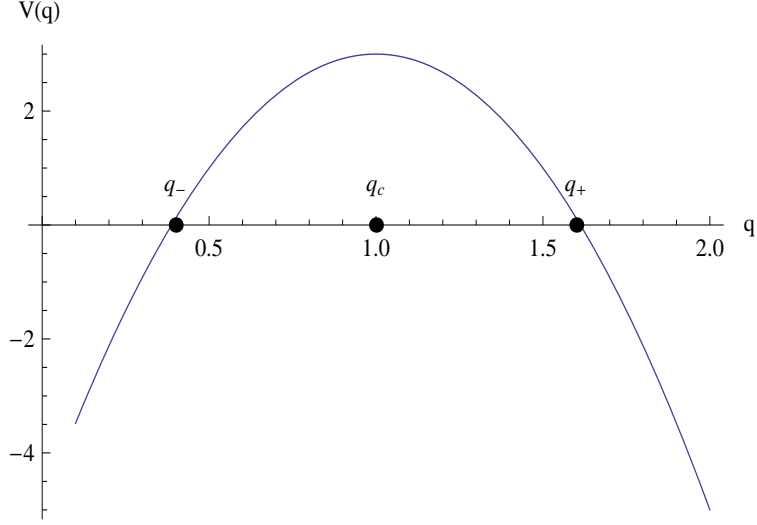}
\caption{Plots of $V_\mathrm{sim}=V(q)$, the simplified model (\ref{Potential-s}), for $q_c=1$, $\mu=8$, and $k=1$.
$q_{\pm}$ denote the zero points of $V(q)$. 
\label{V(q)-simple}}
\end{center}
\end{figure}

By doing this, one can study the model analytically without losing characteristic properties of the original model {within the approximation mentioned above.}
Then we use the following potential $V_\mathrm{sim}$
 (see Fig.~\ref{V(q)-simple}):
\begin{equation}\label{Potential-s}
   V_\mathrm{sim}\equiv 3k- \mu (q_c -q)^2-\beta\ ,
\end{equation}
where $\mu$ and $\beta$ are constant parameters. Hereafter we discuss
 the case of $\beta=0$.\\

Equation of motion for $q$ is easily solved as
\begin{equation}
 q=q_c+a_+e^{\frac{{\cal N}}{t_0}t}+a_-e^{-\frac{{\cal N}}{t_0}t}\, , \quad t_0\equiv \sqrt{{3\over4\mu}}\ .
\end{equation}
The coefficients $a_{\pm}$ are determined through the boundary conditions, 
$q(0)=q_0$ and $q(1)=q_1$:
\begin{eqnarray}
 a_+&=&-{\tilde{q}_0e^{-\frac{{\cal N}}{t_0}}-\tilde{q}_1 \over 2\sinh (\frac{{\cal N}}{t_0})}\, , \\
  a_-&=&{\tilde{q}_0e^{\frac{{\cal N}}{t_0}}-\tilde{q}_1 \over 2\sinh (\frac{{\cal N}}{t_0})}\, , \\
  \tilde{q}_i&\equiv&q_i-q_c\ .
\end{eqnarray}
By substituting these coefficients into (\ref{simple}), we get 
\begin{equation}
 S_\mathrm{sim}^{(0)}=v_3 \mu t_0\left({3k{\cal N}\over \mu t_0}
 -{(\tilde{q}_1^2+\tilde{q}_0^2)\cosh(\frac{{\cal N}}{t_0})-2\tilde{q}_1\tilde{q}_0
                     \over \sinh(\frac{{\cal N}}{t_0})}\right)\ .
\label{eq:simplifiedS0}
\end{equation}
From the stationary condition
$\delta S_\mathrm{sim}^{(0)}/\delta {\cal N}=0$, we find two saddle points,
{say} ${\cal N_{\pm}}$:
\begin{equation}\label{Nsaddle}
   {\cal N_{\pm}}=t_0 \cosh^{-1}\left\{{\mu\over 3k}\left(\tilde{q}_0\tilde{q}_1\pm 
   \sqrt{\left( \tilde{q}_0^2-{3k\over \mu}\right ) \left( \tilde{q}_1^2-{3k\over \mu}\right )} \right)\right\}.
\end{equation}
In order to solve (\ref{Nsaddle}), we parameterize ${\cal N}$ in the polar coordinate 
$(r, \ \theta)$:
\begin{equation}
 e^{\frac{{\cal N}}{t_0}}=re^{i\theta}\, \quad  {\rm or} \quad 
    {\cal N}=(\log r +i~\theta) t_0\ .
\end{equation}
Then (\ref{Nsaddle}) can be rewritten in terms of $r$ and $\theta$:
\begin{eqnarray}
  \cosh \left (\frac{{\cal N}}{t_0}\right )&=&{1\over 2}\left[\left(r+{1\over r}\right) \cos\theta+i\left(r-{1\over r}\right) \sin\theta\right]\, 
                              \label{saddle-3} \nonumber \\
            &=& Q_0Q_1\pm\sqrt{(Q_0^2-1)(Q_1^2-1)}\, \label{saddle-2}\equiv X_{\pm},
\end{eqnarray}
where
\begin{equation}
  Q_0={\tilde{q}_0\over |\tilde{q}_\pm |}\, , \quad Q_1={\tilde{q}_1\over |\tilde{q}_\pm |}\, , 
  \quad \tilde{q}_\pm=\pm \sqrt{{3k\over \mu}}\ .
\end{equation}
Note that $\tilde{q}_\pm$ denote the zeroes of the potential term $V_\mathrm{sim}$ defined by (\ref{Potential-s}).

\vspace{.3cm}
\subsubsection{Tunneling}
\vspace{.3cm}

\begin{figure}[htbp]
\vspace{.3cm}
\begin{center}
\includegraphics[width=10.0cm,height=7cm]{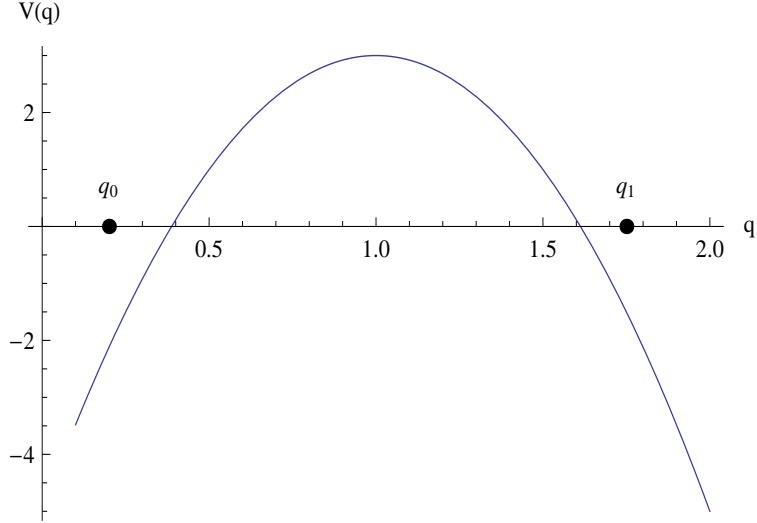}
\caption{Plots of the typical position of $q_0$ and $q_1$ for the tunneling process are shown with
$V_\mathrm{sim}=V(q)$, the simplified model (\ref{Potential-s}) shown in Fig. \ref{V(q)-simple}.
\label{Tunnel-simple}}
\end{center}
\end{figure}

Let us consider a situation that $q_0$ and $q_1$ are put on the opposite side of the hill of
the potential (see Fig.~\ref{Tunnel-simple}). The transition from $q_0$ to $q_1$ would be realized
by the quantum tunneling. 

\begin{figure}[htbp]
\vspace{.3cm}
\begin{center}
\includegraphics[width=10.0cm,height=7cm]{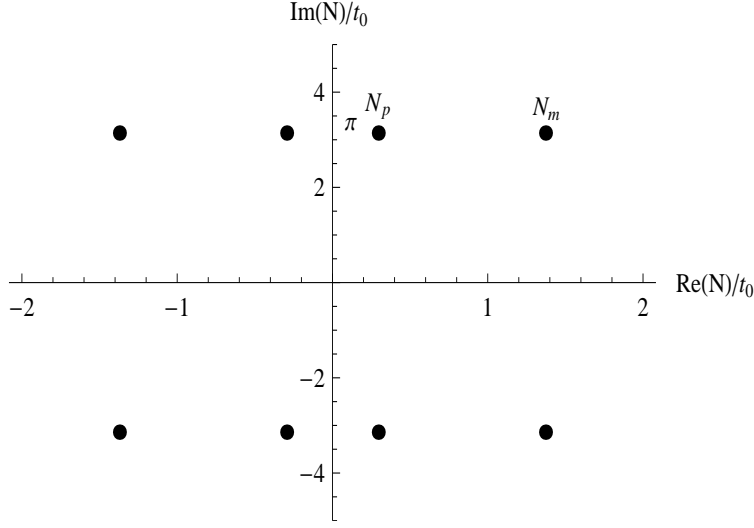}
\caption{Plots of the saddle point of ${\cal N}$ for the simplified model. Here
the case of $\theta=\pi$,  $\tilde{q}_0=-\tilde{q}_+(1+0.2)$ and  $\tilde{q}_1=\tilde{q}_+(1+0.5)$ is shown.
\label{saddle-simple}}
\end{center}
\end{figure}

\vspace{.3cm}
In this case, since $Q_0<-1$ and  $Q_1>1$, $X_{\pm}$ in (\ref{saddle-3}) are real and 
{satisfy} an inequality $X_-<X_+<-1$.
Then we obtain
\begin{eqnarray}
   X_{\pm}&=&-{1\over 2}\left(r_{\pm}+{1\over r_{\pm}}\right) (<-1)\, , \label{saddle-1st} \\ 
       \theta &=& \pm(2n+1)\pi\, , \quad n=0, \pm 1, \pm 2, \ldots
\label{saddle-ph}
\end{eqnarray}
From this, we find two values of $r$, at $r_{\pm l}>1$ and $r_{\pm s}<1$, for $X_{\pm}$.
So there are eight saddle points for one $|\theta|$ in the complex ${\cal N}$ plane.
In Fig.~\ref{saddle-simple}, saddle points for $n=0$ with some specific parameters are shown.

We should notice that there are no real ${\cal N}$ solutions in the
present case.
This implies that the path connecting $q_0$ and $q_1$ represents a quantum process. 


For the saddle points with
$\theta=\pi~~(n=0)$ and $r=r_{\pm s}$, the action is evaluated as
\begin{eqnarray}
  \Re\left[{2\pi^2\over \hbar}iS_0\right]&=&-{2\pi^2\over \hbar}{3\pi t_0}=-{3\sqrt{3}\pi^3\over 
  \sqrt{\mu}\hbar}  \nonumber \\
     &=&-{\sqrt{12}\pi^2\over \hbar}\int_{\tilde{q}_-}^{\tilde{q}_+}\sqrt{3k-\mu\tilde{q}^2}d\tilde{q}=-P(\tilde{q}_+,\tilde{q}_-).
\end{eqnarray}
Then the factor $e^{-P(\tilde{q}_+,\tilde{q}_-)}$ corresponds to the tunneling amplitude.

 As for the above solution, the propagator has a slightly different
phase from the one of $r=r_{\pm s}$. This implies that this solution corresponds to the propagator traveling a slightly long path in the classical region
$0<q<q_-$.

\vspace{.3cm}


\begin{figure}
\begin{center}
\includegraphics[width=162pt]{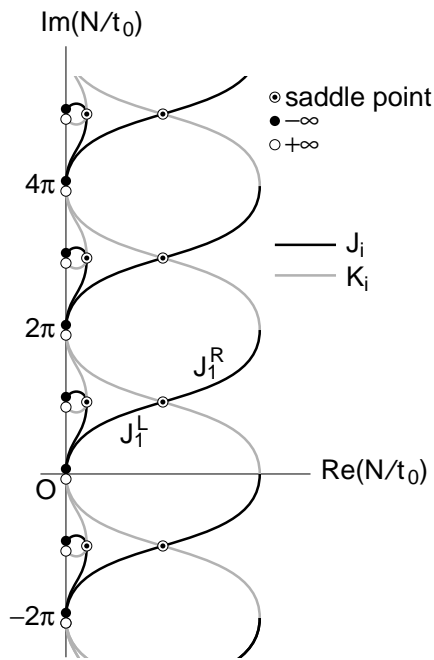}
\caption{
A sketch of the thimbles:
Due to the degeneracy between steepest descent and ascent flows
(thimbles and dual thimbles),
it is not possible to classify the flow segments, definitely.
For each saddle point on the relevant integration path,
a neighboring segment of a steepest descent flow is labeled with $J_i$
and one of a steepest ascent flow is with $K_i$.
}
\label{fig:JKthimbles}
\end{center}
\end{figure}

The propagator is calculated by 
integrating the r. h. s. of (\ref{eq:propagatorG}) over $q$, and
the integration path of $\mathcal{N}$ is deformed to a curve in the first
quadrant of the complex $\mathcal{N}$ plane.
In the method of the Lefschetz thimble, the curved path contains
some steepest descent flows of $\Re[i{S^{(0)}_\mathrm{min}(\mathcal{N})}]$.
As shown in Fig.~\ref{fig:JKthimbles},
there are two possible saddle points each of which connects to the origin
$\mathcal{N}=0$ through a steepest descent flow (a Lefschetz thimble).
One is $\mathcal{N}_p^{(n=0)}=(-\cosh^{-1}(-Q_0)+\cosh^{-1}Q_1+i\pi)t_0$
in the $\mathcal{N}_p$ series and another is
$\mathcal{N}_m^{(n=0)}=(\cosh^{-1}(-Q_0)+\cosh^{-1}Q_1+i\pi)t_0$
in the $\mathcal{N}_m$ series.

\begin{figure}[htbp]
\vspace{.3cm}
\begin{center}
\includegraphics[width=7.0cm,height=7cm]{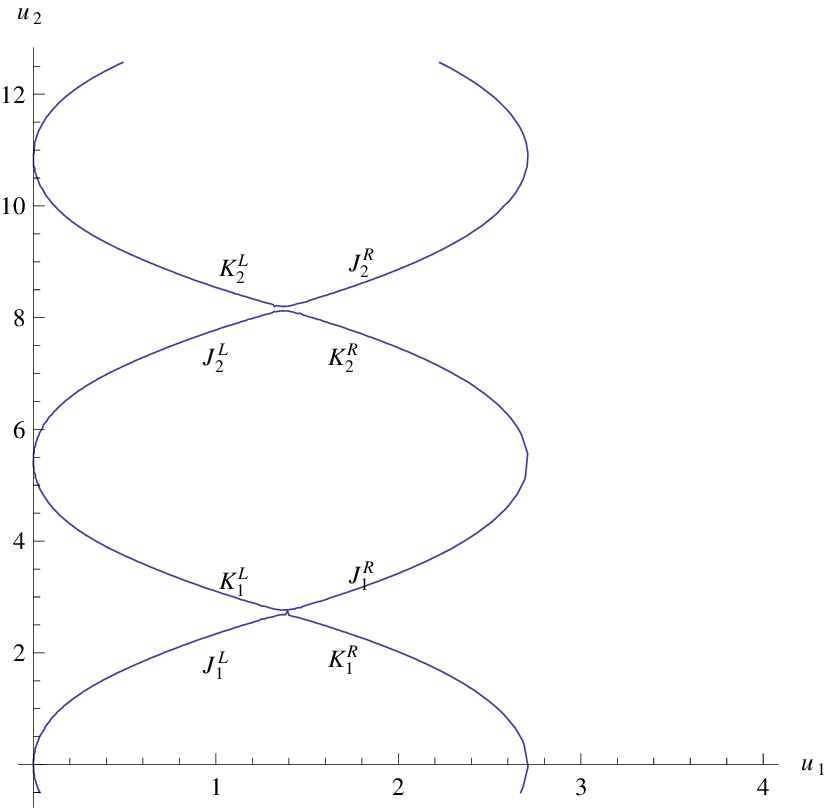}
\includegraphics[width=7.0cm,height=7cm]{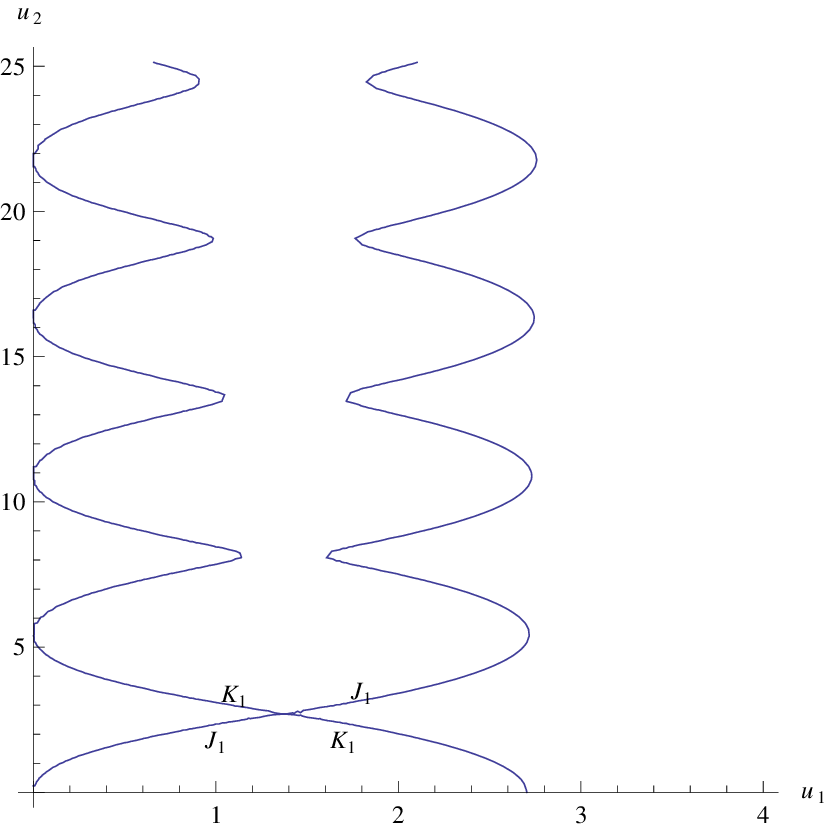}
\caption{Left: Plots of the Lefschetz thimbles for ${\cal N}_m$ of the simplified model. Here
the case of $\theta=\pi$,  $\tilde{q}_0=-\tilde{q}_+(1+0.2)$ and  $\tilde{q}_1=\tilde{q}_+(1+0.5)$ is shown.
Right: The thimbles corresponding to
the first saddle of ${\cal N}_m$ series or an integral path when a perturbation $\Delta S_0=i{\cal N}/100.$
is included.
\label{Thimble-simple}}
\end{center}
\end{figure}

The thimble which passes the former saddle point $\mathcal{N}_p^{(n=0)}$
connects to the singular point at $\mathcal{N}=i(\pi+0)t_0$,
and is terminated there.
This means that the thimble attached to this saddle point is irrelevant to the
present method.
On the other hand, the thimble which passes the latter saddle point
$\mathcal{N}_m^{(n=0)}$ does not reach any singular point before the next
saddle point $\mathcal{N}_m^{(n=1)}=(\cosh^{-1}(-Q_0)+\cosh^{-1}Q_1+3i\pi)t_0$.
This thimble is shown as the set of $J_1^L$ and $J_1^R$
in Fig.~\ref{fig:JKthimbles} and also in the left panel
of  Fig.~\ref{Thimble-simple}.
Furthermore, its dual thimble (the steepest ascent flow) intersects
the real axis of the complex $\mathcal{N}$ plane.
Therefore, the relevant path should run from the origin toward the saddle point
$\mathcal{N}_m^{(n=0)}$.
Now, the Lefschetz thimbles are obtained as the flow lines emanating
from the saddle points of the $\mathcal{N}_m$ series, as shown
in the left of Fig.~\ref{Thimble-simple}.

However, the situation is somewhat complicated because of the degeneracy
between the flow lines $J_i$ and $K_{i+1}$ \footnote{
Here we consider the suffix is identified with  $n=i$.}. 
To find the appropriate path over the saddle point $\mathcal{N}_m^{(n=1)}$,
one may try to provide a perturbation term to the original action
(\ref{eq:simplifiedS0}), for example, $\Delta S_0=i{\cal N}/100$.
Actually, $\Delta S_0$ removes the degeneracy as depicted by the right panel
 of Fig.~\ref{Thimble-simple}. 
Finally, setting $\Delta S_0=0$ again, one finds a unique path as 
$J_1^L \rightarrow J_1^R \rightarrow K_2^R \rightarrow J_2^R \rightarrow \cdots$\ ,
where only the first saddle point $\mathcal{N}_m^{(n=0)}$ dominates
 the integration.




\vspace{.3cm}
\noindent{\bf Tunneling probability via steepest descent method} 

The tunneling probability is estimated by using the propagator (\ref{eq:propagatorG}),
and in the simplified model the amplitude is reduced to
\begin{equation}
G(q_1,q_0)
=\int_{0^+}^\infty\!\!\sqrt{\frac{3i/2t_0}{2\pi\sinh(\frac{{\cal N}}{t_0})}}\,\,
 e^{iS_\mathrm{sim}^{(0)}(\mathcal{N})}\,d\mathcal{N}\label{eq:G10(N)}
\end{equation}
after integrating over $q$.

As mentioned above, 
 $G(q_1,q_0)$ is approximately
 evaluated by the main contribution of the saddle point at
\begin{equation}
\mathcal{N}_0\equiv
\mathcal{N}_m^{(n=0)}=(\cosh^{-1}(-Q_0)+\cosh^{-1}Q_1+i\pi)\,t_0\ ,
\end{equation}
which is the joint of $J_1^L$ and $J_1^R$.
Then, the amplitude (\ref{eq:G10(N)}) is calculated as follows:
\begin{eqnarray}
G(q_1,q_0)
&\simeq&\sqrt{\frac{3i/2t_0}{2\pi\sinh(\frac{{\cal N}}{t_0})}}\,\,
\,e^{i\Re S_\mathrm{sim}^{(0)}(\mathcal{N}_0)}
\!\!\mathop{\int}_{J_1^L+J_1^R}\!\!\!
e^{-\Im S_\mathrm{sim}^{(0)}(\mathcal{N})}\,d\mathcal{N}
\\
&\simeq&\sqrt{\frac{3i/2t_0}{2\pi\sinh(\frac{{\cal N}}{t_0})}}\,
e^{i\Re S_\mathrm{sim}^{(0)}(\mathcal{N}_0)}
\!\!\int_{-\infty}^\infty\!\!\!\!
e^{-\Im S_\mathrm{sim}^{(0)}(\mathcal{N}_0)-\frac{1}{2}|S_\mathrm{sim}^{(0)}{}''(\mathcal{N}_0)|\nu^2}
e^{i\theta_0}d\nu \\
&=&\sqrt{\frac{3i/2t_0}{2\pi\sinh(\frac{{\cal N}}{t_0})}}\,
e^{i(\Re S_\mathrm{sim}^{(0)}(\mathcal{N}_0)+\theta_0)}
e^{-\Im S_\mathrm{sim}^{(0)}(\mathcal{N}_0)}\,
\sqrt{\frac{2\pi}{|S_\mathrm{sim}^{(0)}{}''(\mathcal{N}_0)|}}\ ,
\end{eqnarray}
where $\theta_0$ is the incident angle of the thimble $J_1^L$ into the point
$\mathcal{N}_0$ in the complex $\mathcal{N}$ plane.
The angle $\theta_0$ is determined by
\begin{equation}
\tan\theta_0
=\frac{\Re S_\mathrm{sim}^{(0)}{}''(\mathcal{N}_0)}{|S_\mathrm{sim}^{(0)}{}''(\mathcal{N}_0)|+\Im S_\mathrm{sim}^{(0)}{}''(\mathcal{N}_0)}
=\frac{|S_\mathrm{sim}^{(0)}{}''(\mathcal{N}_0)|-\Im S_\mathrm{sim}^{(0)}{}''(\mathcal{N}_0)}{\Re S_\mathrm{sim}^{(0)}{}''(\mathcal{N}_0)}\ .
\end{equation}

Since it holds that
\begin{equation}
S_\mathrm{sim}^{(0)}{}''(\mathcal{N})
=\frac{3v_3k}{t_0}\,F^{(2)}\left ( \frac{{\cal N}}{t_0} \right )
\end{equation}
with
\begin{equation}
F^{(2)}\left ( \frac{{\cal N}}{t_0} \right )
=\left[\frac{2Q_0Q_1\cosh^2(\frac{{\cal N}}{t_0})
-2(Q_0^2+Q_1^2)\cosh(\frac{{\cal N}}{t_0})+2Q_0Q_1}{\sinh^3(\frac{{\cal N}}{t_0})}\right]
\ ,
\end{equation}
the second order derivative $S_\mathrm{sim}^{(0)}{}''(\mathcal{N})$ satisfies that
\begin{equation}
\Re S_\mathrm{sim}^{(0)}{}''(\mathcal{N}_0)>0\ ,\quad
\Im S_\mathrm{sim}^{(0)}{}''(\mathcal{N}_0)=0\ .
\end{equation}
Therefore, the angle $\theta_0$ is found to  be $\frac{\pi}{4}$.

Now, the absolute value of the tunneling amplitude is given by
\begin{eqnarray}
|G(q_1,q_0)|
&=&\sqrt{\frac{3/2t_0}{|\sinh(\frac{{\cal N}}{t_0})\,S_\mathrm{sim}^{(0)}{}''(\mathcal{N}_0)|}}\,
e^{-\Im S_\mathrm{sim}^{(0)}(\mathcal{N}_0)}\\
&=&\sqrt{\frac{1}{2v_3k|\sinh(\frac{{\cal N}}{t_0})\,F^{(2)}(\frac{{\cal N}}{t_0})|}}
\,e^{-\Im S_\mathrm{sim}^{(0)}(\mathcal{N}_0)}\ ,
\label{eq:absG}
\end{eqnarray}
in which the factor $|\sinh(\mathcal{N}_0/t_0)\,F^{(2)}(\mathcal{N}_0/t_0)|$ is a certain function of $Q_0$ and $Q_1$, and
\begin{equation}
-\Im S_\mathrm{sim}^{(0)}(\mathcal{N}_0)=-3kv_3t_0\pi\ . 
\end{equation}

We can see $\Im S_\mathrm{sim}^{(0)}(\mathcal{N}_0)=P(q_+,q_-)$,
where $P(q_+,q_-)$ is defined by (\ref{WDW-pot-1}) in Sec.~\ref{sec:QCos-2}
when $V$ is replaced by the simplified potential $V_{\rm sim}$ used
in {the present} section. This implies the result obtained
here reproduces the tunneling probability $e^{-2P(q_+,q_-)}$ given
 in Sec.~\ref{sec:QCos-2}.
From Eq. (\ref{eq:absG}), it is found that $|G(q_1,q_0)|^2$ depends on $t_0$ such as
\begin{equation}
|G(q_1,q_0)|^2\propto e^{-6kv_3t_0\pi}\ .
\end{equation}

\vspace{.3cm}
\subsubsection{Saddles for periodic Euclidean solution}

In the previous subsection, we considered the saddle-point contribution from the $n=0$ sector.
Then a natural question arises: What is the meaning of other infinite series of the 
saddle points for $n\geq 1$? {We could answer to the question by considering the ``instanton''\cite{Barv}. 
In \cite{Barv}, the authors have found a periodic Euclidean solution of the Friedmann equation 
regarded as an instanton one, which oscillates between $q_-$ and $q_+$.}

In the following, 
let us estimate the instanton contribution by using the Euclidean path-integral formalism.
In our simplified model with $V_\mathrm{sim}$, equations of motion for $q$ in the Euclidean time $\tau$ 
is given by
\begin{equation}
  \left( {\partial q \over \partial \tau}\right)^2=-{4\over 3} \mu (q-q_c)^2+4k
\end{equation}
Then the solution is easily obtained as
\begin{equation}
  q_I = \sqrt{{3\over \mu}} \sin \left( \sqrt{{ 4\mu  \over 3}}\right) +q_0
      = 2t_0 \sin \left(\frac{\tau}{t_0}\right) +q_0
\end{equation}
The period of the solution, $T_0$, becomes
\begin{equation}
   T_0=2\pi t_0=\pi\sqrt{\frac{3}{\mu}}\, .
\end{equation} 
Then we get the Euclidean action for the
instanton solution:
\begin{eqnarray}
  S_0^{E} &=&i v_3\int_0^{T_0} d\tau \left({3\over 4} \dot{q}^2+3k-\mu (q-q_c)^2 \right)\,   \\
   &=& iv_3\times 3T_0\ .
\end{eqnarray}
Therefore, we find
\begin{equation}
  \Re\left[{2\pi^2\over \hbar}\,iS_0^{E}\right]
 = -{2\pi^2\over \hbar}{6\pi t_0}=-{6\sqrt{3}\pi^3\over \sqrt{\mu}\hbar}
 =-2 P(\tilde{q}_+,\tilde{q}_-)\ .
\end{equation}

\vspace{.3cm}
On the other hand, 
for the saddle point solutions corresponding to (\ref{saddle-ph}), we have 
\begin{equation}
  \Re\left[{2\pi^2\over \hbar}\,iS_0\right] = -(2n+1) P(\tilde{q}_+,\tilde{q}_-)\, .
\end{equation}
From this result, we could say that the saddle contribution with $n \geq 1$ represents  
that from ``$n$ instantons''.
They are exponentially suppressed compared to the leading $n=0$ contribution.
{These saddle points for $n\geq 1$, however, do not contribute as saddles in the integral given above
to evaluate the propagator due to the path which is chosen as the Lefschetz thimble of $n=0$ saddle point.}


\section{Summary and Discussions}

In this paper, cosmology driven by SYM theory is studied in the FRW space-time. 
Here all the SYM fields are integrated out and the vacuum expectation value of 
its energy-momentum tensor $\langle T_{\mu\nu}^{\mathrm{SYM}}\rangle$ is given by holographic method.
Based on the equations of motion with this $\langle T_{\mu\nu}^{\mathrm{SYM}}\rangle$, we introduced
a new simple effective action which could reproduce the equations of motion of the theory.
 
The SYM theory provides two kinds of terms
in the action, loop correction term and a radiation term.
The gravitational part includes the
4D cosmological constant $\Lambda_4$.
The value of $\Lambda_4$ is restricted to be positive in order to realize 
the inflationary universe at large scale factor $a_0$.
In the present case, however, it is not completely free.
When SYM theory is included, $\Lambda_4$ is bounded from above, so the expansion rate 
at large $a_0$ is modified by the loop correction of the SYM fields. 
 
In the region of small $a_0$, the radiation plays an important role. 
Although the magnitude of the radiation is arbitrary,
there appears a lower bound of $a_0$. 
Further, a 
small scale universe with the radiation could be born and appear at large $a_0$ after a quantum tunneling. 
Here, this phenomenon is studied through two quantum cosmological methods for mini-superspace of gravity. 

One is to solve the WDW equation.
Here the WDW equation is easily given by considering our effective action introduced as mentioned above.
In this sense, our effective action is very useful to study the quantum mechanics of the theory.
The tunneling probability is calculated by imposing an appropriate boundary condition
at large $a_0$ for both cases of $C=0$ and $C>0$.

The other is to calculate the same quantity by Lorentzian path-integral according to the method proposed
by \cite{Turok}. This method is useful when the radiation is absent. The result coincides with the solution of
the WDW equation. 

On the other hand, it is difficult to proceed the calculation in terms of the effective action
used for the WDW equation when the radiation exists. So we consider a simplified model in which the
potential in the WDW equation is approximated by the harmonic form. 
In this case, we could show a Lefschetz thimble as the unique path
 for the tunneling propagator. The calculation
is performed according to the steepest descent method, and the result coincides with the solution of the WDW equation
with the same harmonic potential. We should notice
that, for the harmonic potential, we have many saddle points in the complex lapse ${\cal N}$ plane. However,
the path, which contributes to the Lorentzian path-integral, has only one saddle which corresponds to the
tunneling.

Although, here, we concentrated to the tunneling amplitude, there are other kinds of propagators whose
initial and final values of $a_0$ are different from the one of the tunneling case. The situation is also depending on
the form of potential. It is characterized by the radiation and $\Lambda_4$.
On such various kinds of propagators, we will discuss in the future.

\vspace{.3cm}


\section*{Acknowledgments}
One of the authors (M.T.) is supported in part by the JSPS Grant-in-Aid
 for Scientific Research, Grant No.\,16K05357,
 and he is grateful for helpful discussion with Yuya Tanizaki.


\newpage

\appendix

\noindent{\bf\Large Appendix}

\section{The Wheeler-DeWitt equation}
\label{sec:WDWequation}

At first, we give the Wheeler-DeWitt equation used to obtain the wave-function of the universe.
\vspace{.5cm}
(\ref{4d-action-q4}) is written as
\begin{equation}
  S=\int\!\!dt\,L
\end{equation}
\begin{equation}
L=\mathcal{N}a_0^3\left(\frac{3}{a_0^2}\left(-\frac{\dot{a}_0^2}{\mathcal{N}^2}+
k\right)-\Lambda_{\mathrm{eff}}\right)v_3\, , \quad v_3={V_3 \over \kappa_4^2}
\end{equation}
where
\begin{equation}
\Lambda_{\mathrm{eff}}=3\lambda_-=3\,
\frac{~~1-\sqrt{1-4\tilde{\alpha}^2\left(
\frac{\Lambda_4}{3}+\tilde{\alpha}^2\frac{4}{R^2}\frac{C}{a_0^4}\right)}~~}{2\tilde{\alpha}^2}
\end{equation}

\begin{equation}
p_a=\frac{\partial L}{\partial\dot{a}_0}
=-{6\over\mathcal{N}}a_0\dot{a}_0 v_3
\end{equation}
Then
\begin{eqnarray}
H
&=&p_a\dot{a}_0-L\\
&=&-\frac{\mathcal{N}}{12a_0 v_3}\left(
p_a^2+{12a_0^4 v_3^2}\left(\frac{3}{a_0^2}k-\Lambda_{\mathrm{eff}}\right)\right)\\
&=&\mathcal{N}\hat{H}
\end{eqnarray}
where 
\begin{equation}
  \hat{H}=-\frac{1}{12a_0 v_3}\left(
p_a^2+{12a_0^4 v_3^2}\left(\frac{3}{a_0^2}k-\Lambda_{\mathrm{eff}}\right)\right)
\end{equation}
Then we have
\begin{equation}
  S=\int dt ~(p_a\dot{a}_0-\mathcal{N}\hat{H})
\end{equation}

This indicates the lapse function $\mathcal{N}$ is a Lagrange multiplier providing the constraint
\begin{equation}
\hat{H}=0
\end{equation}
This is written to a quantized form by using
\begin{equation}
p_a\rightarrow -i\frac{\partial}{\partial a_0}
\end{equation}
as
\begin{equation}\label{WDW-eq}
\left(-\frac{\partial^2}{\partial a_0^2}
+{12a_0^2v_3^2} \left(
3k-a_0^2 \Lambda_{\mathrm{eff}}\right)\right)\,\Psi(a_0)=0
\end{equation}
This equation could be applied to the region where a classical solution for $a_0(t)$ is forbidden.
In the below, by changing the variable in this equation, we show some numerical result for the tunneling process under an appropriate
boundary conditions.


\section{Derivation of Eq. (\ref{effL-2})}
\label{sec:derivation}

{The} result (\ref{effL-2}) is obtained as follows. Assuming the form of $L_{\mathrm{SYM}}^{\mathrm{eff}}$ as,
\begin{equation}\label{Assume}
   L_{\mathrm{SYM}}^{\mathrm{eff}}=h_0(a_0)+{\dot{a}_0^2\over {\cal N}^2} h_2(a_0)+{\dot{a}_0^4\over {\cal N}^4} h_4(a_0)\, .
\end{equation}
This form would be supported by the analyticity and $g_{0i}=0$ gauge (\ref{RW-N}). By substituting this into (\ref{4d-action-q2}),
we find from $\delta S/\delta {\cal N}=0$,
\begin{equation}
    \left\{{1\over \kappa_4^2}
      \left(3 a_0({\dot{a}_0^2\over {\cal N}^2}+k)-a_0^3
      \Lambda_4\right)+ a_0^3\left(h_0-{\dot{a}_0^2\over {\cal N}^2} h_2-3{\dot{a}_0^4\over {\cal N}^4} h_4\right) 
         \right\} \delta{\cal N}=0
\end{equation}
This is compared with (\ref{bc-RW2}) for ${\cal N}=1$ gauge,
\begin{equation}\label{bc-RW21}
 \lambda\equiv  \left({\dot{a}_0\over a_0}\right)^2+{k\over a_0^2} = {\Lambda_4\over 3}+{\kappa_4^2\over 3} 
\langle T_{00}^{\mathrm{SYM}}\rangle\, ,
\end{equation}
where 
\begin{equation}
  T_{00}^{\mathrm{SYM}}=\rho
={3\alpha\over 16}\left(\frac{4C}{R^2a_0^4}+\lambda^2\right)\, ,
 \quad \lambda={\dot{a}_0^2\over a_0^2}+{k\over a_0^2}
\end{equation}

Then we have,
\begin{eqnarray}
  h_0 &=&-{3\alpha\over 16}\left(\frac{4C}{R^2a_0^4}+{k^2\over a_0^4}\right)\, , \\
   h_2 &=& {3\alpha\over 8} { k\over a_0^4}\, , \\
   h_4 &=& {\alpha\over 16} { 1\over a_0^4}\, ,
\end{eqnarray}
Plugging this into (\ref{Assume}) we find (\ref{effL-2}), which is expressed for general  ${\cal N}$.

\section{Picard-Lefschetz method}
\label{Picard-Lefschetz}

In this Appendix, let us introduce the Picard-Lefschetz method
\footnote{The notation of this Appendix owes to \cite{Tanizaki-Tachibana}.}.
In the path-integral formalism, the partition function is defined by 
\be
Z=\int_{\mathcal{M}} d\Phi \exp\left({\cal F}(\Phi)/\hbar \right), 
\ee
where $\mathcal{M}$ is the target space of parameters $\Phi$. 
Since this integral is, in general,  an multi-dimensional oscillatory integral, 
we need a technique to consider about it. Here we use the Lefschetz-thimble method 
\cite{Witten}\cite{sign}.

The basic idea is to deform the integration contour $\mathcal{M}$ into steepest descent cycles inside its complexified space $\mathcal{M}_{\mathbb{C}}$ by using the Cauchy theorem when ${\cal F}$ is holomorphic. 
We denote the holomorphic coordinate of $\mathcal{M}_{\mathbb{C}}$ as $\Phi=(z^1,\ldots,z^n)$, 
and the set of saddle points as 
\be
\Sigma=\{z_{\sigma}\}:=\left\{\frac{\partial {\cal F}}{\partial z^i}=0 \right\}. 
\ee
Using the K\"ahler metric on $\mathcal{M}_\mathbb{C}$, $ ds^2=g_{i\overline{j}}dz^i\otimes  d\overline{z^j}$, we define the gradient flow by 
\be
\frac{d z^i}{d t}=g^{i\overline{j}}{\left(\frac{\partial \overline{{\cal F}}}{\partial \overline{z^j}}\right)}. 
\label{eq:gradient_flow_general}
\ee
As an important property of this differential equation, we have
\be
{d{\cal F}\over dt}=\left|\partial {\cal F}\right|^2\ge 0. 
\ee
Therefore, along the flow line, the real part of the free energy increases while its imaginary part stays constant. This means that we can define the steepest descent and ascent cycles associated with each saddle point $z_{\sigma}$ by this gradient flow. Using solutions of the gradient flow $z(t)$, they are defined as 
\be
\mathcal{J}_{\sigma}=\{z(0)\, |\, z(t)\to z_{\sigma}, t\to -\infty\}, \quad 
\mathcal{K}_{\sigma}=\{z(0)\, |\, z(t)\to z_{\sigma}, t\to +\infty\}. 
\ee
These are called Lefschetz thimbles and dual thimbles. 
They are dual quantities in terms of the intersection pairing $\langle\cdot,\cdot \rangle$, i.e., $\langle \mathcal{J}_{\sigma},\mathcal{K}_{\tau}\rangle =\delta_{\sigma \tau}$, which means that one can decompose $\mathcal{M}$ in terms of $\mathcal{J}_{\sigma}$ as 
\be
\int_{\mathcal{M}} d \Phi \exp\left({\cal F}(\Phi)/\hbar\right)=\sum_{\sigma\in \Sigma} \langle  \mathcal{M}, \mathcal{K}_{\sigma}\rangle \int_{\mathcal{J}_{\sigma}} d^n z \exp\left({\cal F}(\Phi)/\hbar\right). 
\ee
If all $\mathrm{Re}({\cal F}(\Phi_{\sigma}))$ are different with each other in the limit $\hbar\to 0$, we replace the integral by the saddle-point approximation. Then we obtain at the leading order that 
\be
Z=\sum_{\sigma} \langle  \mathcal{M}, \mathcal{K}_{\sigma}\rangle  \exp\left({\cal F}(z_{\sigma})/\hbar\right). 
\label{eq:MFpartition_function_complex_saddles}
\ee

We can summarize the necessary steps of the mean-field approximation with the sign problem as follows: 
\begin{enumerate}
\item Complexify the target space $\mathcal{M}$ to $\mathcal{M}_{\mathbb{C}}$, and find the saddle points $z_{\sigma}$ by solving the equation $\partial {\cal F}=0$ in $\mathcal{M}_{\mathbb{C}}$. 
\item Solve the gradient flow (\ref{eq:gradient_flow_general}), and construct Lefschetz thimbles $\mathcal{J}_{\sigma}$ and dual thimbles $\mathcal{K}_{\sigma}$.
\item Pick up the saddle point $z_{\sigma}$ that has the minimal free energy $\mathrm{Re}({\cal F}(z_{\sigma}))$ with nonzero intersection number $\langle\mathcal{M},\mathcal{K}_{\sigma}\rangle$.  
\end{enumerate}



\end{document}